\documentclass[10pt,letterpaper]{article}

\usepackage{cogsci}

\cogscifinalcopy 

\usepackage{color}
\usepackage{pslatex}
\usepackage{apacite}
\usepackage{float} 
\usepackage{graphicx}
\usepackage{subcaption}
\usepackage[labelformat=simple]{subcaption}

\usepackage{enumitem}

\title{Bayesian Inference of Social Norms \\ as Shared Constraints on Behavior}
 
\author{{\large \bf Zhi-Xuan Tan (xuan@mit.edu)} 
  \and {\large \bf Desmond C. Ong (desmond.c.ong@gmail.com)} \\
  A*STAR Artificial Intelligence Initiative,
Agency for Science, Technology and Research \\
1 Fusionopolis Way, Singapore 138632}

\begin{document}

\maketitle

\begin{abstract}

People act upon their desires, but often, also act in adherence to implicit social norms. How do people infer these unstated social norms from others' behavior, especially in novel social contexts? We propose that laypeople have intuitive theories of social norms as behavioral constraints shared \emph{across} different agents in the same social context. We formalize inference of norms using a Bayesian Theory of Mind approach, and show that this computational approach provides excellent predictions of how people infer norms in two scenarios. Our results suggest that people separate the influence of norms and individual desires on others' actions, and have implications for modelling generalizations of hidden causes of behavior. 

\end{abstract}

\section{Introduction}

Imagine entering a cafeteria in a foreign country that you know little about. There are but a handful of individuals in the cafeteria; you notice a tray-return receptacle at the far end, but you do not notice any signage on the walls detailing the expectations governing tray returns. If you observe someone carry their tray to the far end in order to return it, what inferences might you make? Does that person \emph{like} returning trays, incurring the cost of walking across the room to do so? Or, is there an implicit social norm at play? A second person leaves without returning their tray. What might you now infer about them or about the potential social norm? Lastly, a third person approaches the second and loudly chastises them for not returning their tray, what would you then infer about everybody's preferences and the social factors at play?

Social norms are ubiquitous features of human societies, and as the example above suggests, we are able to rapidly infer their presence in novel situations. Children as young as three \cite{schmidt2016young} demonstrate the ability to learn and generalize these rules of social behavior \cite{rakoczy2013early}. Not surprisingly, this ability continues into adulthood, allowing us, for example, to travel to new countries and then learn through observing others whether one is obliged to tip at restaurants, what the appropriate manner of greeting is, or what topics of conversation are considered impolite. In other words, we seem to possess not just an intuitive Theory of Mind (ToM) which allows us to infer the beliefs and desires of individuals, but also a corresponding ability to efficiently make inferences about shared norms that drive behavior \emph{across} individuals. Furthermore, we appear naturally capable of disentangling the influence of social norms and individual desires: when we see someone picking up trash on the sidewalk, we infer that this is more likely due to an obligation to keep the streets clean rather than enjoyment of the act itself.

Despite (or perhaps because of) its ubiquity, how people \emph{infer} social normativity is relatively understudied. The philosophical literature on social norms has generally focused on characterizing the precise nature of such norms---whether they are best understood as social practices, preferences conditioned upon shared expectations of behavior, or commonly-held normative attitudes \cite{bicchieri2005grammar, brennan2013explaining}. Across philosophy, economics and psychology, there has also been an emphasis upon understanding the conditions and mechanisms for the emergence of norms \cite{hawkins2018emergence}---whether they arise, for example, as Nash equilibria \cite{axelrod1986evolutionary,young2015evolution}, correlated equilibria \cite{gintis2010social}, or through maximization of cultural values \cite{boloni2018towards}.
Other studies investigate how norms influence decision making \cite{chang2011great}, and how they are enforced \cite{fehr2004third}. 
However, apart from a few simulation-based studies \cite{savarimuthu2010obligation,cranefield2016bayesian}, research into how social norms are \emph{inferred} remains scarce.

How then to explain our ability to infer social norms? In recent years, Bayesian models of cognition have begun to establish a computational basis for how people make social inferences. The Bayesian Theory of Mind (BToM) approach is perhaps the most prominent example, allowing researchers to formalize how people make graded judgments about unobservable mental states by observing the actions of others \cite{baker2017rational}. This approach to social cognition has been extended to model reasoning about others' emotions \cite{ong2015affective}, inferring others' beliefs and desires from observed actions and emotional expressions \cite{wu2018rational}, reasoning about how others balance costs and rewards in deciding how to act \cite{jara2016naive}, learning how people value the welfare of others \cite{kleiman2017learning}, and inferring the presence of co-operation or competition from the behavior of multiple individuals \cite{shum2019theory}. Related work modelling human concept learning as Bayesian rule inference \cite{goodman2008rational} has also been used to develop theories of why people tend to learn act-based moral rules rather than outcome-based ones \cite{nichols2016rational}.

We build upon this tradition of computational cognitive modelling, and hypothesize that people intuitively understand social norms as factors of behavioral influence which are shared across agents in a particular social context. These shared norms influence behavior alongside the individual desires of agents, and can generally be understood as injunctions or constraints on behavior, i.e., they prescribe, recommend, or prohibit certain kinds of actions\footnote{Here we are not interested in modelling descriptive norms---statistical regularities---since they can be directly learned through observation, though it is an interesting but separate research question as to how people might infer injunctive norms from descriptive ones.}. We propose that people include these norms in their lay theories of social behavior, and we model these theories as Bayesian networks which include both norms and desires as possible causes of action. Judgments about the presence of a norm can thus be modelled using Bayesian inference conditioned upon observed actions, which can be made alongside desire inferences. Furthermore, since social norms are shared, inferences about them can be made from observations of multiple agents, unlike those for desires. We discuss several of these models, each of which captures a plausible intuitive theory of how norms influence both desires and actions. We then describe an experiment to test which model provides the best explanation of lay people's judgments in two social scenarios.

\section{Computational Models}

In order to study how people make inferences about social norms given observations of behavior, we choose to model situations where norm-driven actions are likely to be salient. In such situations, agents can take the role of \textbf{actors}, who are in the position to comply with a potentially applicable norm, or they can take the role of \textbf{judges}, who are in the position to enforce that norm after observing non-compliance. For simplicity, we restrict ourselves to the smallest multi-agent setting, with only one actor and one judge. The actor takes an action $A_1$, which corresponds to compliance or non-compliance with a potential norm. We also assume that the actor has some latent (binary) desire $D_1$ over action $A_1$ and its associated outcome. If the actor decides not to comply with the norm, the judge may take an action $A_2$, which corresponds to enforcement or non-enforcement of the potential norm. The judge also has a (binary) desire $D_2$ over the space of outcomes of $A_1$ (note however that $D_2$ is conditionally independent of $A_1$, since the judge's desires exist whether or not $A_1$ is taken). $D_2$ influences the enforcement action $A_2$ because $A_2$ can rectify the outcome produced by $A_1$.  We denote the norm by $N$, which either exists ($N=1$) or does not ($N=0$) in the modelled situation.

\begin{figure}[t]
    \centering
    \includegraphics[width=0.90\linewidth]{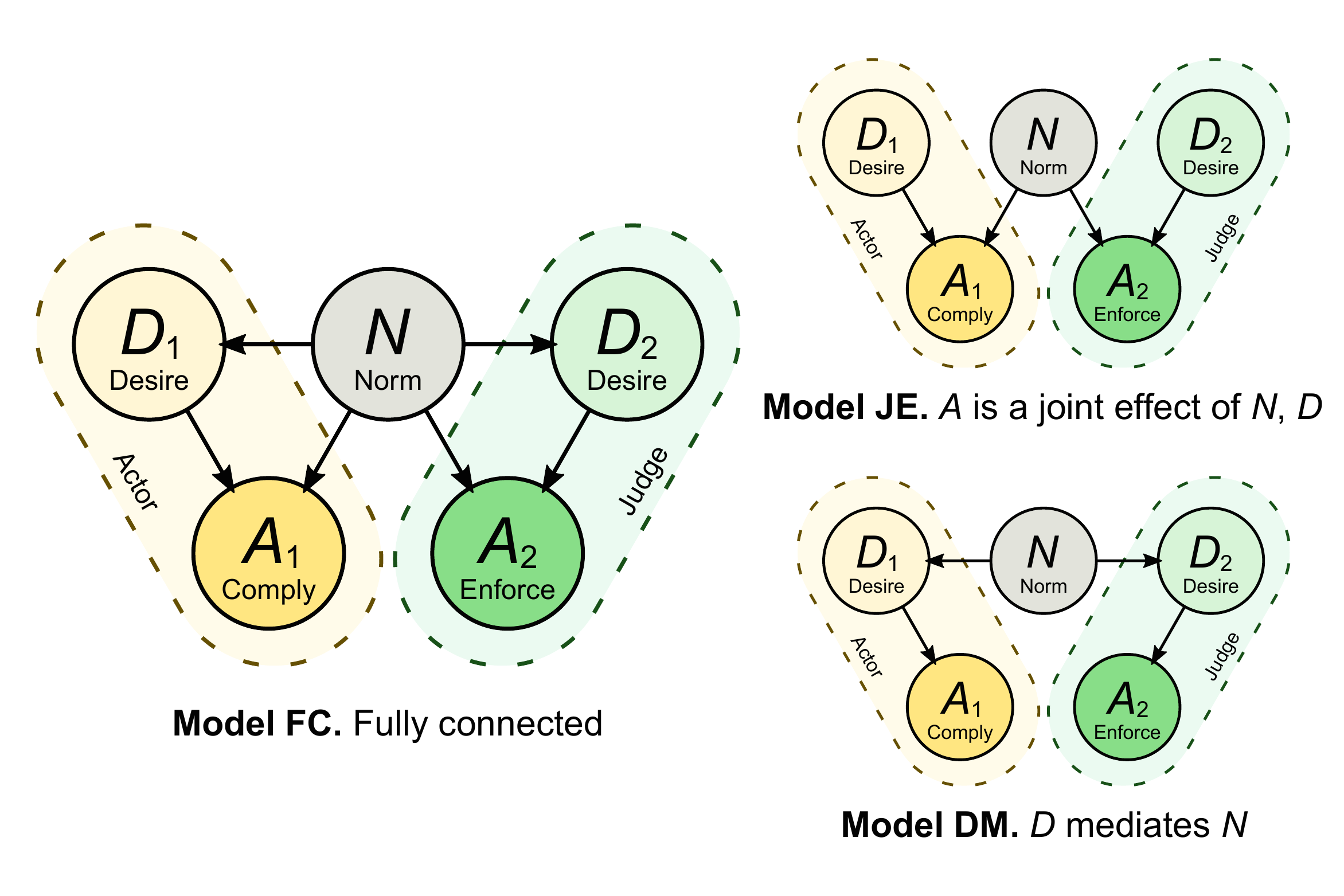}
    \caption{Possible intuitive causal models of how social norms influence behavior. We consider two agents, an Actor (on the left of each model, with nodes in yellow) and a Judge (on the right, with nodes in green). In Model FC, norms influence desires, and both norms and desires jointly influence actions. In Model JE, desires are independent of norms, while in Model DM, desires are the sole mediators of norms upon behavior.}
    \label{fig:models}
\end{figure}

We consider three plausible models of laypeople's intuitive theories of norm-driven behavior (Fig. \ref{fig:models}). In the \textbf{Fully Connected (FC)} model, we posit that social norms influence the desires of each agent, reflecting the idea that norms can be (partly) internalized, such that norm compliance becomes part of one's desire. We also assume that norms and desires jointly affect an agent's actions. This corresponds to the notion that whether or not a norm is internalized, it continues to directly influence actions, whether by imposing an expected social cost to non-compliance, or simply by having a normative force that is separate of individual desire. 

In our second candidate model, the \textbf{Joint Effect (JE)} model, desires and norms still jointly influence actions, but agent desires are independent of norms. That is, lay people's notions of what agents `want' to do are separate from what they `should' do, mapping roughly to the Kantian distinction between desire and duty. Lastly, the \textbf{Desire Mediation (DM)} model assumes that norms do not influence actions directly, but only when mediated by desires. This assumption corresponds to the Humean notion that desires are the sole motives for action---that `shoulds' cannot influence action unless they also become `wants'---a notion which might plausibly feature in lay intuitions about norms. In addition to these three `complete' models, we also tested two lesioned models: a desire-only model (\textbf{D-only}) and a norm-only model (\textbf{N-only}).

It is worth emphasizing that in proposing these models, we are \emph{not} attempting to give a rigorous philosophical account of the relationship between norms, desires, and actions, nor are we attempting to argue that social norms cannot be explained in terms of desires, or for that matter, other cognitive variables such as shared beliefs or expectations. Rather, our intention is to propose models of how people \emph{intuitively} understand norms and norm-driven behavior.

\section{Experiment}

We conducted an experiment through Amazon's Mechanical Turk (AMT) to elicit lay people's likelihood judgments about norms, desires and actions in two different social scenarios: one involving an obligative norm---the norm that people should return their trays after eating---and another involving a prohibitive norm---the norm that people should not litter. Given widespread intuitive acceptance of the act-omission distinction \cite{kahneman2005cognitive}, we expected that people might also respond differently to obligations and prohibitions. We also chose common, but not universal, social norms, so that we could better observe how people adjusted their certainty about the existence of each norm.
\subsection{Methods}

We provide a sample of our experiment, our data, and analysis code at \url{\footnotesize https://github.com/ztangent/norms-cogsci19}.

\subsubsection{Scenario Structure.}

Both scenarios were identical in structure---participants were introduced to an actor in a position to comply with a potentially applicable norm, and asked to make various likelihood judgments. They were then shown the actor not complying with the potential norm, introduced to a judge who (unknown to the actor) had observed the actor's action, and made another round of judgments.

\subsubsection{Experimental Conditions.}

To measure how well our proposed models predict lay people's inferences, as well as determine which model best captures intuitions about norms, we divided each each scenario into five conditions, each querying for different sets of likelihood judgments:

\begin{enumerate}[label=\Alph*.,noitemsep]
    \item Norm and desire priors, and desires given norms: \\
    $P(N)$, $P(D_1)$, $P(D_2)$, $P(D_1|N)$, $P(D_2|N)$.
    \item Actions conditioned on desires only: $P(A_1|D_1)$, $P(A_2|D_2)$.
    \item Actions conditioned on norms only: $P(A_1|N)$, $P(A_2|N)$.
    \item Actions conditioned on both: $P(A_1|D_1,N)$, $P(A_2|D_2,N)$.
    \item Norm and desire posteriors: \\ $P(D_1|A_1)$, $P(D_2|A_1,A_2)$, $P(N|A_1)$, $P(N|A_1,A_2)$
\end{enumerate}

Data from conditions A--D were used both to calibrate the models and to investigate people's intuitions about the relationship between norms, desires and actions, e.g., by comparing $P(D_1|N)$ (condition B) to $P(D_1)$ (condition A) to see if people judge desires to be dependent upon norms. After calibrating the models, data from condition E was compared against the models' posterior inferences to see if they predicted participants' inferences about norms and desires.

\subsubsection{Participants.}
We recruited $200$ US participants (mean age 35.6, SD 11.2; 104 male, 95 female, 1 unreported) via AMT, restricting to those with a HIT approval rate of $99\%$ and above. All participants went through both scenarios in random order, and each participant was randomly assigned to a different condition within each scenario (i.e. assigned conditions for Scenario 1 and 2 were independent).  For Scenario 1, conditions 1A through 1E, sample sizes were $n=$51, 24, 25, 51 and 49 respectively. For Scenario 2, conditions 2A through 2E, sample sizes were $n=$49, 25, 25, 50 and 51 respectively. Assuming a large effect size (Cohen's $f=0.5$), these sample sizes give $>93\%$ power for one-way ANOVA between sub-conditions at the $5\%$ significance level (e.g. comparing $P(A_1|N\!=\!1)$ and $P(A_1|N\!=\!0)$ in condition C).

\subsection{Scenario 1: An Obligative Norm}

\begin{figure}[htb]
    \centering
    \begin{subfigure}[b]{0.4\linewidth}
    \centering
    \includegraphics[width=\textwidth]{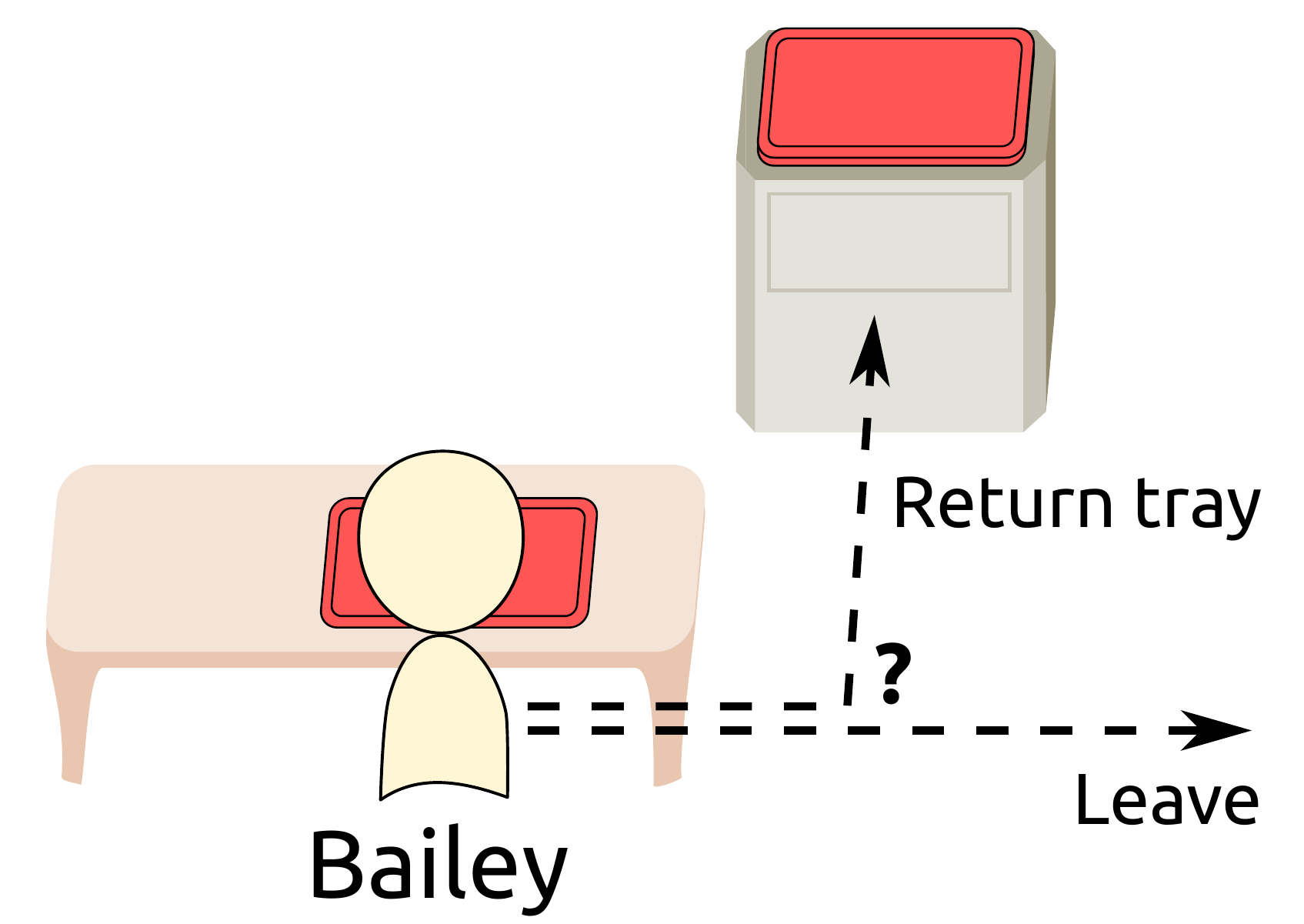}
    \caption{Actor's choices}
    \label{fig:scenario1actor}
    \end{subfigure}
    \hspace{0.1\linewidth}
    \begin{subfigure}[b]{0.4\linewidth}
    \centering
    \includegraphics[width=\textwidth]{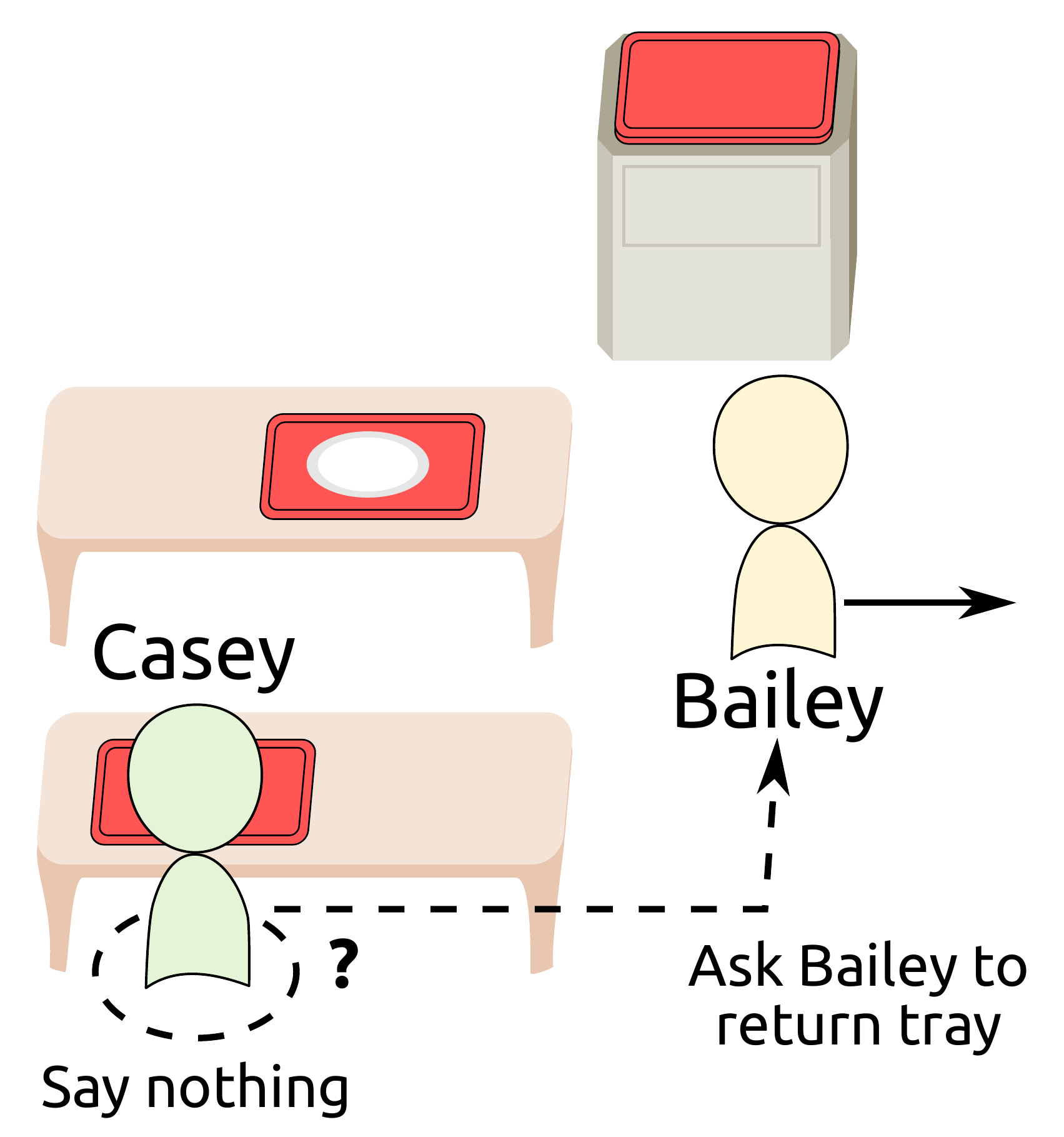}
    \caption{Judge's choices}
    \label{fig:scenario1judge}
    \end{subfigure}
    \caption{Returning one's tray as an obligative social norm.}
    \label{fig:scenario1vignette}
\end{figure}

\begin{figure*}[t!]
    \centering
    \includegraphics[width=0.90\textwidth]{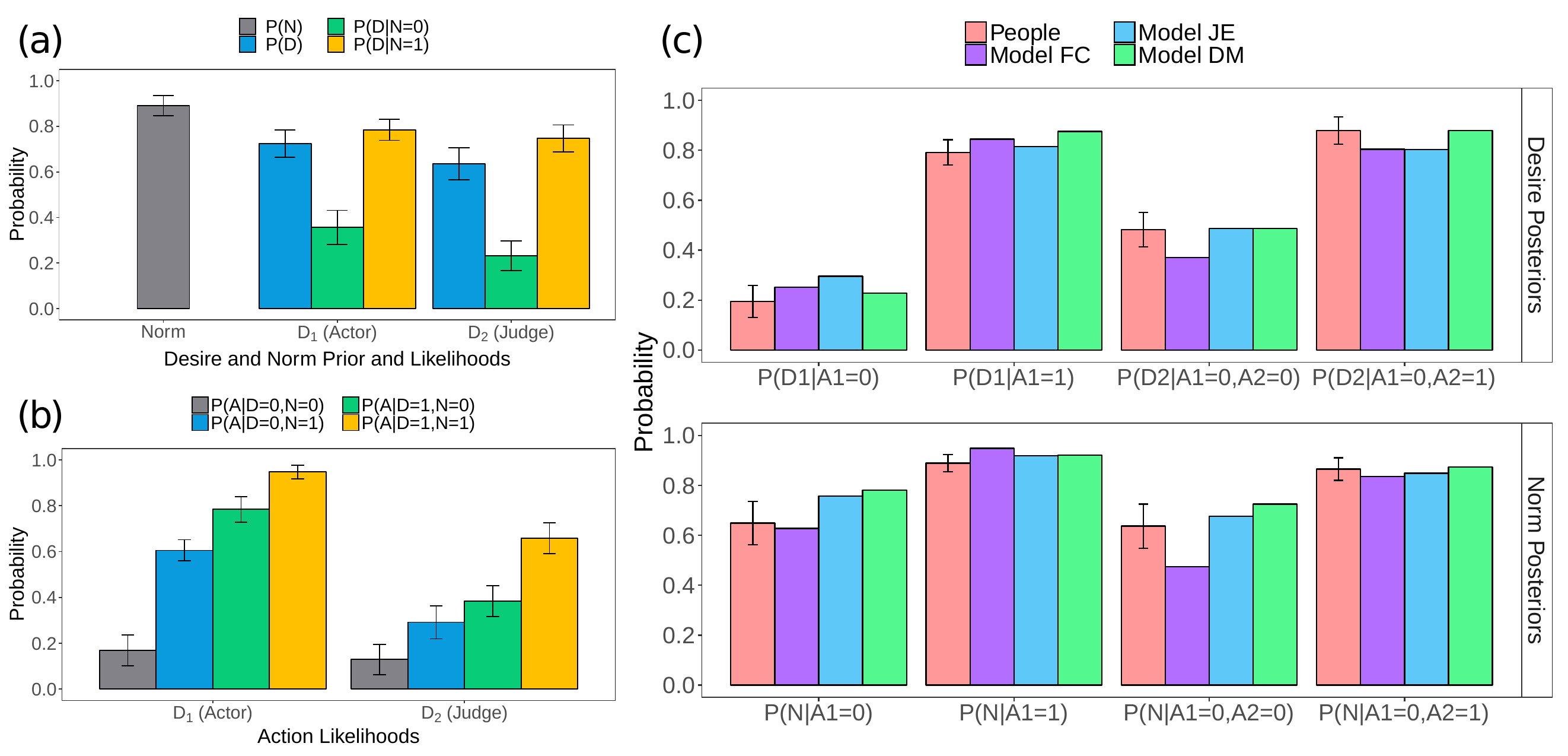}
    \caption{Scenario 1 Results. (a) Empirical norm and desire judgments from condition 1A. (b) Empirical action likelihoods, from condition 1D. (c) Comparing empirical posteriors from condition 1E with posterior judgments from our proposed models.}
    \label{fig:scenario1}
\end{figure*}

\subsubsection{Details.}

Participants were presented with a vignette with two phases, with text descriptions accompanied by illustrations. In the first phase (Figure \ref{fig:scenario1actor}), Bailey, the actor, has finished a meal served on a tray in a restaurant. The restaurant has a tray return station, and Bailey can choose to either return ($A_1\!=\!1$) or leave ($A_1\!=\!0$) the tray. Since the norm at play was obligative ('People should return their trays after eating.'), $A_1\!=\!1$ corresponds to compliance with the norm (if it exists, i.e, if $N\!=\!1$). In the second phase, participants are told that Bailey decides to leave the tray, and are introduced to Casey, who, unknown to Bailey, has watched this occur (Figure \ref{fig:scenario1judge}). Casey now has the option of either asking Bailey to return the tray ($A_2\!=\!1$) or saying nothing ($A_2\!=\!0$), where $A_2\!=\!1$ corresponds to enforcement of the potential norm.

Questions were presented after \emph{each} phase was introduced, asking participants to make judgments depending on the condition they were assigned. When queried for priors (condition A), participants were directly asked how likely they thought a certain state of affairs was true (e.g. "How likely do you think Casey \emph{wants} the tray to be returned?" for $P(D_2)$, "How likely do you think the following norm exists?" for $P(N)$). When queried for conditional likelihoods, including the posteriors, participants were first asked to suppose a certain state of affairs (e.g. "If you saw Casey ask Bailey to return the tray," for $P(\cdot|A_1\!=\!0,A_2\!=\!1)$  or "Suppose that Bailey does not want to return the tray." for $P(\cdot|D_1\!=0)$), and then asked to give likelihood judgments given those suppositions. To make these counterfactuals more concrete in the case of the posterior inferences (condition E), we also provided corresponding illustrations of the counterfactual actions.

\subsubsection{Results.}

To determine if participants intuitively judged desires to be independent of norms, we first analyzed the data from condition A to see if $P(D_i)$, $P(D_i|N\!=\!1)$ and $P(D_i|N\!=\!0)$ ($i\in\{1,2\}$) exhibited significant differences. As can be seen in \ref{fig:scenario1}(a), this was indeed the case, with one-way ANOVA giving $F(2,150)\!=\!55.13$, $p<0.001$, for the actor's conditional and prior desires ($D_1$), and $F(2,150)\!=\!66.95$, $p<0.001$, for the judge's ($D_2$) conditional and prior desires. This provides evidence against Model JE, which assumes that desires are independent of norms.

Next, to determine if desires and norms jointly influence behavior, we analyzed the conditional action likelihoods from condition D. As Figure \ref{fig:scenario1}(b) shows, given a fixed value of desire, there were significant differences between action likelihoods when the norm was either absent or present (all $t$s $>5.77$, all $p$s $<0.001$, df$\!=\!50$, paired test). That is, regardless of whether the agent \emph{wanted} to act, the presence of the tray-return norm ($N\!=\!1$) led people to judge both norm compliance ($A_1\!=\!1$) and norm enforcement ($A_2\!=\!1$) as more likely. This provides evidence against Model DM, which assumes norms have no direct effect on actions. (For brevity, we omit comparisons between $P(A_i|D_i,N)$, $P(A_i|D_i)$ and $P(A_i|N)$, $i \in \{1,2\}$ using the data from conditions B and C, but these display significant differences as well.)

Finally, we computed the desire and norm posteriors under the FC, JE and DM models, then compared them against participants' posterior judgements, as shown in Figure \ref{fig:scenario1}(c). All three models displayed high correlation with the empirical data (FC: $r\!=\!0.944$, JE: $r\!=\!0.974$, DM: $r\!=\!0.981$). The correlations of the lesioned models were worse by comparison (D-only: $r\!=\!0.944$, N-only: $r\!=\!0.384$). Both norm and desire posteriors increased when compliance ($A_1\!=\!1$) or enforcement ($A_2\!=\!1)$ were observed, and decreased otherwise. The three models also captured people's ability to integrate information across multiple agents to infer the presence of norms---when non-compliance by the actor ($A_1\!=\!0$) is observed, the likelihood of the norm's existence decreases ($P(N)>P(N|A_1\!=\!0)$), but when enforcement by the judge ($A_2$) is subsequently observed, the likelihood of the norm increases again ($P(N|A_1\!=\!0)<P(N|A_1\!=0,A_2\!=\!1)$).

Despite the desire and action likelihoods providing strong evidence against the JE and DM models, these models were surprisingly more correlated with participants' posterior judgements than the FC model. The results for Scenario 1 are thus hard to interpret conclusively. Plausibly, this was due the high degree of inter-subject variance in likelihood ratings, suggesting that people's intuitive models of social normativity have substantial heterogeneity.

\subsection{Scenario 2: A Prohibitive Norm}

\begin{figure}[htb]
    \centering
    \begin{subfigure}[b]{0.4\linewidth}
    \centering
    \includegraphics[width=\textwidth]{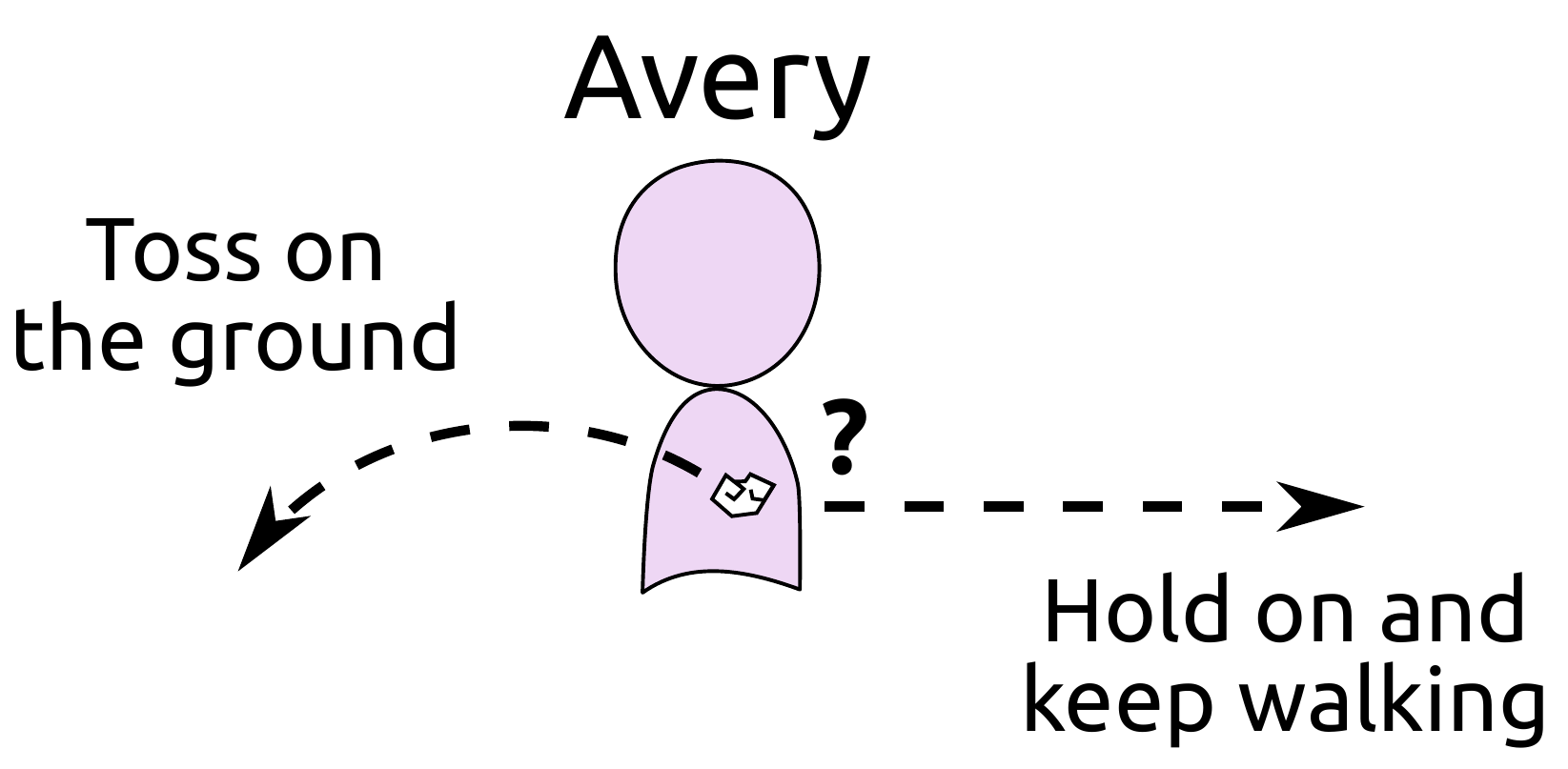}
    \caption{Actor's choices}
    \label{fig:scenario2actor}
    \end{subfigure}
    \hspace{0.1\linewidth}
    \begin{subfigure}[b]{0.4\linewidth}
    \centering
    \includegraphics[width=\textwidth]{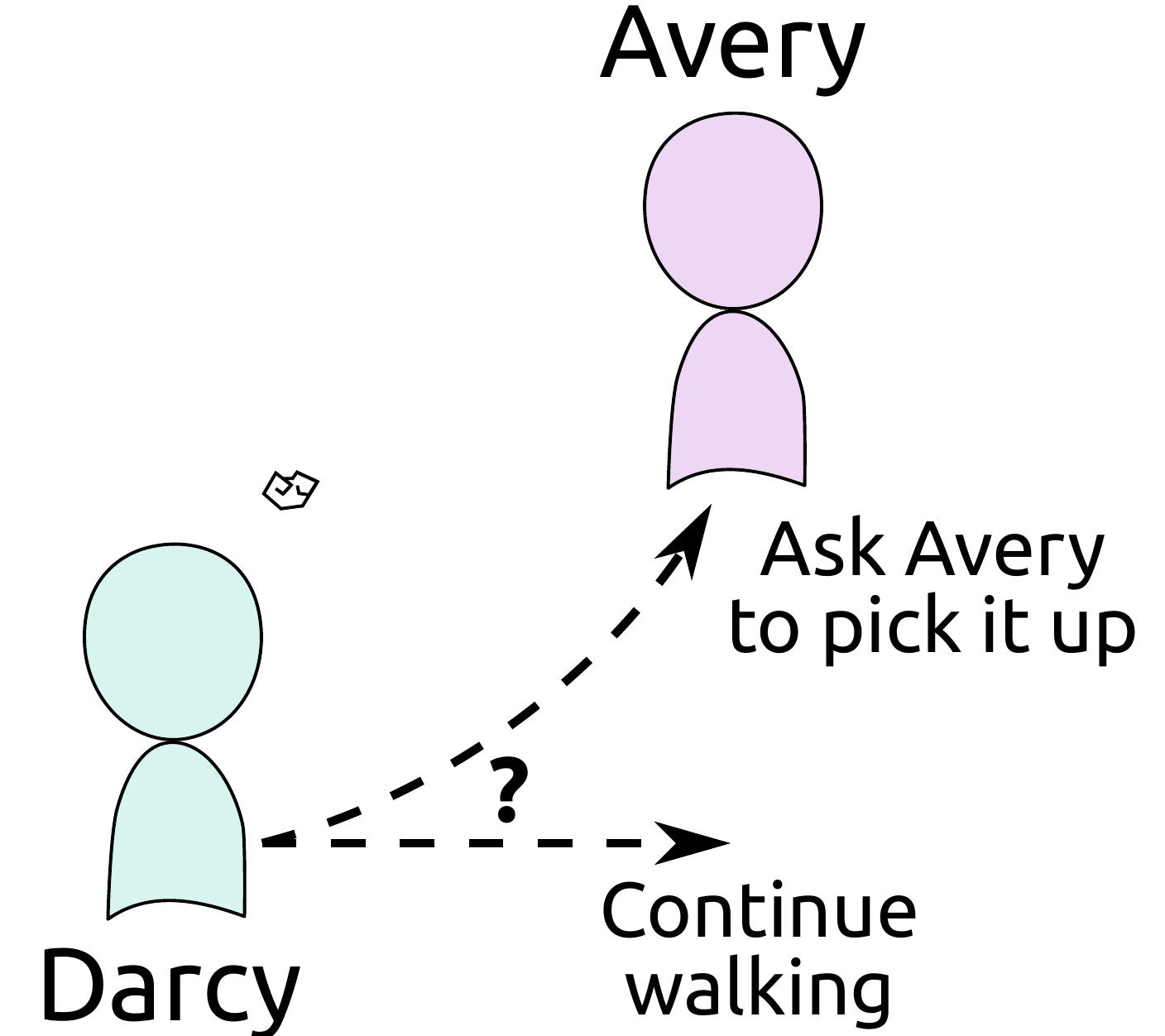}
    \caption{Judge's choices}
    \label{fig:scenario2judge}
    \end{subfigure}
    \caption{Not littering as a prohibitive social norm.}
    \label{fig:scenario2vignette}
\end{figure}

\begin{figure*}[t!]
    \centering
    \includegraphics[width=0.90\textwidth]{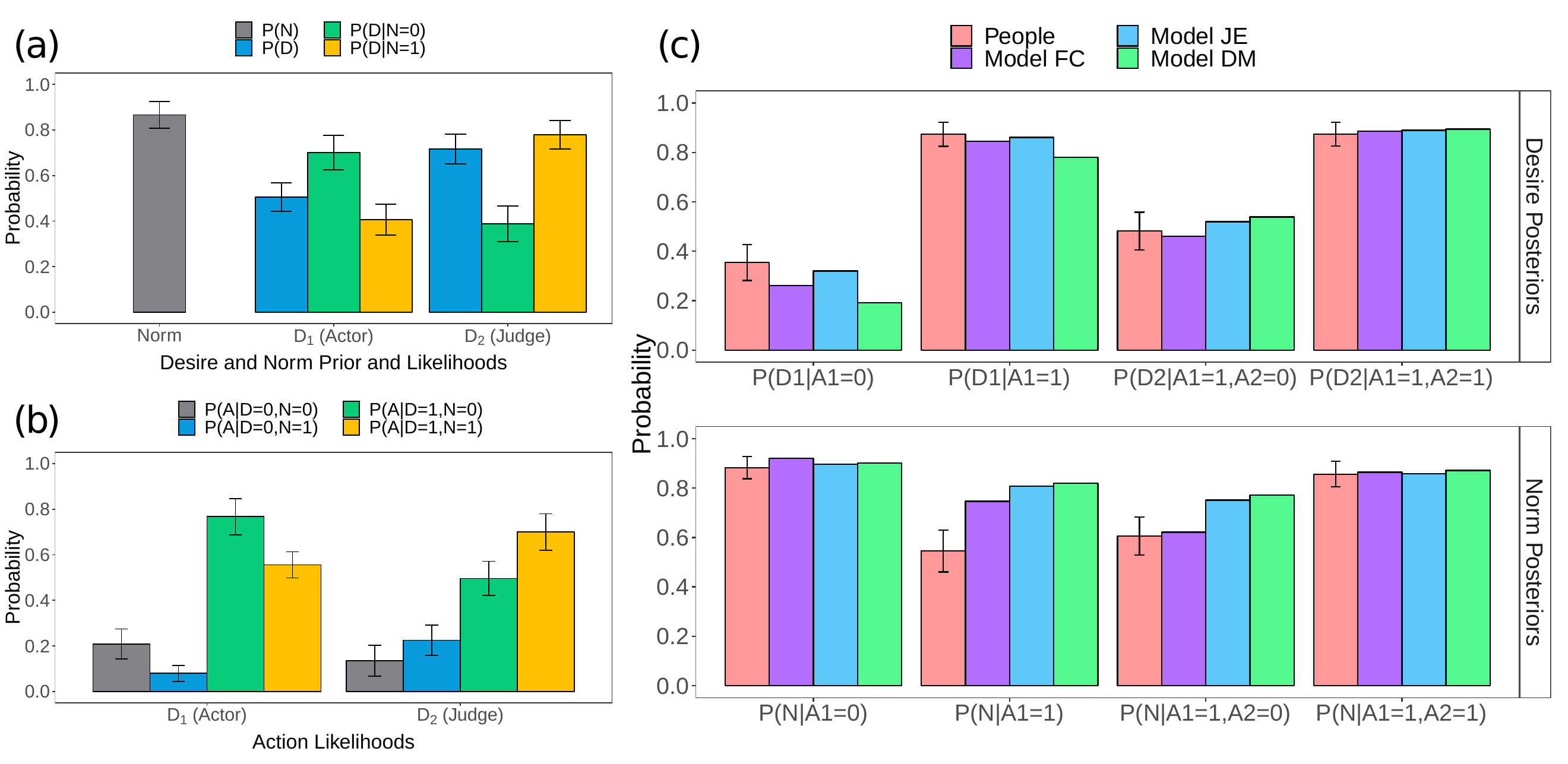}
    \caption{Scenario 2 Results. Figure is similar to Fig. \ref{fig:scenario1} except for the fact that Scenario 2 involved a prohibitive norm, where $A_1=1$ (littering) contravenes the norm. This explains the different patterns for the likelihood of Actor variables ($D_1$, $A_1$).}
    \label{fig:scenario2}
\end{figure*}

\subsubsection{Details.}

As in Scenario 1, participants were presented with a two-phase vignette. In the first phase (Figure \ref{fig:scenario2actor}), Avery, the actor, is walking along a city street while holding on to some crumpled paper. Avery can choose to either toss the paper ($A_1\!=\!1$) or continue holding on ($A_1\!=\!0$). Since the norm at play was prohibitive ('People should not discard their belongings on the ground.'), $A_1\!=\!1$ corresponds to violation of the norm, unlike in Scenario 1. In the second phase, participants are told that Avery decides to toss the crumpled paper on the ground, and are introduced to Darcy, who, unknown to Avery, has watched this occur (Figure \ref{fig:scenario2judge}). Darcy now has the option of either asking Avery to pick up the crumpled paper ($A_2\!=\!1$) or saying nothing ($A_2\!=\!0$), where again $A_2\!=\!1$ corresponds to enforcement of the potential norm. After each phase was introduced, participants were asked to make likelihood judgments depending on the condition they were assigned, using the same question formats as in Scenario 1. 

\subsubsection{Results.}

First, we analyzed the data from condition A to see if participants judged desires to be norm-dependent. As \ref{fig:scenario2}(a) shows, this was once more the case, with one-way ANOVA giving $F(2,144)\!=\!18.02$, $p<0.001$, for the actor's conditional and prior desires ($D_1$), and $F(2,144)\!=\!35.86$, $p<0.001$, for the judge's ($D_2$) conditional and prior desires. We then analyzed the conditional action likelihoods from condition D, and found similarly that there were significant differences when the norm was either absent or present (all $t$s $>3.10$, all $p$s $<0.005$, df$\!=\!49$, paired test). To be clear, the influence of the norm here was in the \emph{opposite} direction---$N\!=\!1$ led to littering ($A_1\!=\!1$) being less likely.

Lastly, we computed the desire and norm posteriors under the various models and compared against the empirical data, as shown in Figure \ref{fig:scenario2}(c). Compared to Scenario 1, more pronounced differences could be observed between models. In particular, Model DM significantly over-estimates the norm's likelihood when a norm-violating or non-enforcing action is taken (see Figure \ref{fig:scenario2}(c) $P(N|A_1\!=\!1)$, $P(N|A_1\!=1,A_2\!=\!0)$), because it attributes the cause of the action primarily to the desire not to comply with or enforce the norm. In contrast, Model FC better predicts that the norm's likelihood should decrease when norm violation is observed. This is because there are multiple causal pathways that lead from the norm to the actions in Model FC---norm-violating actions directly imply that the norm is unlikely to exist, but they also imply that the desire for the norm-violating action exists, which indirectly implies the non-existence of the norm. Model JE over-estimates the norm's likelihood slightly less than Model DM, but still more than Model FC, because it has only one causal pathway from the norm to the action.

Nonetheless, all three models still correlated highly with the data (FC: $r\!=\!0.934$, JE: $r\!=\!0.887$, DM: $r\!=\!0.828$), and the lesioned models again performed worse (D-only: $r\!=\!0.772$, N-only: $r\!=\!0.461$). Both norm and desire likelihoods increased when compliance ($A_1\!=\!0$) or enforcement ($A_2\!=\!1)$ were observed, and decreased otherwise. We similarly observed that people integrated information from multiple agents: the likelihood of the norm decreases after observing norm violation ($P(N|A_1\!=\!1)<P(N)$), but increases again after subsequently observing enforcement ($P(N|A_1\!=1,A_2\!=\!1)>P(N|A_1\!=\!1)$).

Unlike in Scenario 1, Model FC displayed the highest degree of correlation out of our three proposed models. This, combined with the analysis showing that both actions and desires are directly norm-dependent, provide strong evidence for model FC over the other two models. One reason this might have been the case for Scenario 2 is that a prohibitive norm tends to \emph{conflict} with the direction of desire---people are more likely to act how they believe they \emph{should}, whatever they happen to \emph{want} for themselves. This would disfavor Model DM (hence its over-estimation of the norm's likelihood) but favor Model FC, because it betters captures the restraining effect of norms on both desires and actions.

\section{General Discussion}

We experimentally investigated how laypeople infer norms from behavior, and showed that a Bayesian model provides excellent predictions of people's posterior inferences of both obligative and prohibitive norms. We tested several plausible theories, and found strong evidence that people understand norms to directly influence both desires and actions. This suggested that the JE and DM models should be ruled out, leaving the FC model. While our analysis of the model's posterior inferences in Scenario 1 did not unambiguously support this conclusion, the corresponding analysis for Scenario 2 did so, with the FC model showing the highest correlation when averaged across both scenarios (FC: $r\!=\!0.939$, JE: $r\!=\!0.931$, DM: $r\!=\!0.905$). Furthermore, comparison with lesioned models showed that accurate inferences cannot be made by omitting either desires on norms.

Our results lend support to our hypothesis that people understand social norms as behavioral constraints shared \emph{across} agents, in contrast to preferences that are idiosyncratic to individual agents. In this way, people are able to observe different actions made by individuals in different roles, integrating that information and allowing them to rapidly make inferences about the presence of social norms in a given context. This ability is highly useful, for it allows us navigate unfamiliar social environments without deducing the preferences of every stranger about how one should act.

Of course, not all environments are unfamiliar---people spend their whole lives with a familiar set of norms. Thus, one might expect a person to bring strong expectations to bear when making inferences about norms in a new, but familiar, situation. Indeed, this was the case for many participants in our experiment---while we constructed scenarios with common but not universal norms, participants were often \emph{certain} that the norm in question was present, even before observing any actions. These priors made it harder for our experiment to detect whether people deemed a norm \emph{more} likely to exist after observing norm compliance, but made it easier to detect when people deemed a norm less likely to exist after observing a norm  violation. Future experiments should introduce participants to a more alien environment where they have no sense of what the norms might be, and see if they can infer the presence of a norm through a few observations.

Participants not only came in with varied priors, but also varied likelihood judgements, with some participants giving more weight to the influence of norms than others. This heterogeneity may help explain why our results for Scenario 1 were not conclusive---participants' internal models might have diverse parameters, and some might even have different model structures altogether. As such, while the results clearly showed that the models predicted average judgments with high correlation, the ability to distinguish the exact model type may have been lost. Still, it is interesting that even with such diversity, people are still able to rapidly learn and converge upon the same set of norms. How exactly this convergence occurs is a topic worth exploring further.

In conclusion, we have demonstrated a principled, computational framework for how people infer the shared drivers of behavior that we call social norms. In addition, our results give us insight into people's intuitive theories of norms as influencing \emph{both} our desires and our actions. This builds upon previously studied models that infer the beliefs, desires, and intentions of single agents, extending their inferential capacity to large groups of agents constrained by shared context. By laying the groundwork for how we make these inferences, our work elucidates one way in which we make sense of the full richness of social life---and how we \emph{ought} to live it.

\section{Acknowledgments}

This work was supported by the A*STAR Human-Centric Artificial Intelligence Programme (SERC SSF Project No. A1718g0048).

\bibliographystyle{apacite}

\setlength{\bibleftmargin}{.125in}
\setlength{\bibindent}{-\bibleftmargin}
\bibliography{biblio}

\end{document}